\documentstyle[preprint,aps]{revtex}
\tightenlines

\def\ra{\rightarrow}

\def\be{\begin{equation}}
\def\ee{\end{equation}}
\def\bea{\begin{eqnarray}}
\def\eea{\end{eqnarray}}

\begin{document}
\preprint{\vbox{\baselineskip=14pt
%\begin{flushright}
\rightline{UH511-889-97}  \break
\rightline{March 1999} %\break
%\rightline{hep-ph/9903332}
  }}
%\end{flushright}
\vspace{.25in}
\title{The $\rho - \pi$ Puzzle of $J/\psi$ and $\psi'$ Decays
\vspace{.25 in} \\
March 23, 1999} 
\author{S.F. TUAN} 
\address{Department of Physics, 2505 Correa Road \\
University of Hawaii at Manoa \\
Honolulu, HI  96822, USA%}
%\author{ A.N. OTHER }
\\
\vspace{.25in}
[Submitted to Commun. Theor. Phys.]}

%\address{Department
%of Physics, Theoretical Physics, 1 Keble Road,\\
%Oxford OX1 3NP, England}
%%%%%%%%%%%%%%%%%%%%%%%%%%%%%%%%
%
% You may repeat \author \address
% as often as necessary
%
%%%%%%%%%%%%%%%%%%%%%%%%%%%%%%%
\maketitle

\abstract{The recent BES Collaboration data on $\psi' \ra PV$, 
particularly the isospin violating
mode $\psi' \ra \pi^0 \omega^0$ and finding of a finite number for 
$B(\psi' \ra K^{*0}\overline{K}^0)$, 
enable us now to deal more precisely about the challenges to
theory concerning this extraordinary and remarkable so called $\rho - \pi$
Puzzle of $J/\psi$ and $\psi'$ Decays.  
In terms of existing data, and deploying the simplest phenomenology, 
measurement of $\psi' \ra \pi^+\pi^-$ and whether a finite number 
for the $K^{*+} K^-$ mode
might require a significantly larger accumulation of data, remain
interesting questions.  
\vspace{.50in}

     The strong suppression of $\psi ' \ra \rho \pi$ and $K^*\overline{K}$ 
decays is well known and has been the
subject for considerable experimental and theoretical activity.  No evidence 
is seen for $\psi' \ra \rho \pi$ in the BES group's 3.8 million  
$\psi'$ event sample, in striking contrast to the case for 
the $J/\psi$, where $J/\psi \ra \rho \pi$ 
is the dominant decay.  This conundrum
is referred to as the ``$\rho \pi$" puzzle and remains one of the most 
intriguing mysteries in quarkonium physics. The BES group$^{\cite{olsen}}$ 
has the largest event sample of $\psi' $ decays
thus affording it the opportunity to undertake a systematic investigation of 
$\psi'$ decays to all of the lowest lying vector- plus pseudoscalar-meson 
(VP) final states.
Of particular interest is the identification of the isospin violating mode
$\psi' \ra \pi^0 \omega^0$ and the finding of a finite value for the 
branching ratio $B(\psi' \ra K^{* 0} \overline{K}^0$) which is likely an
isospin violating VP case.  These new numbers impel us by creating a ``crisis''
to deal in a systematic manner both theoretically and experimentally to
pin down the essential nature of this
so called $\rho-\pi$ Puzzle of $J/\psi$ and $\psi'$ decays.  
Indeed interesting further but pedestrian tests appear feasible for the BES 
sample within a phenomenological framework, 
e.g. measurement of $\psi' \ra \pi^+ \pi^-, \omega \eta,
\rho \eta'$ and $\rho \eta,$
though to reach for $\psi'$ to the charged $K^{*+}K^-$ VP mode might 
require a significantly larger data sample such as may be possible 
with the construction of a new main drift chamber MDC III and a future 
tau-charm factory at IHEP.

     On the theoretical side, there has also been much recent 
activity$^{\cite{li}}$.
Li, Bugg, and Zou$^{\cite{li}}$ argued that final-state interactions involving 
the rescattering of $a_1 \rho$ and $a_2$ into 
$\rho \pi$ could be important and might interfere destructively 
in the case of $\psi'$.  The possibility of a destructive interference
in $\psi'$, though it is fortuitous, cannot be ruled out in view
of Suzuki's finding$^{\cite{li}}$ of the large long distance final-state interaction
in
$J/\psi$ based on {\underline{existing}} data. However, this
interference model does appear to have more assumptions than
predictions!  Brodsky and Karliner$^{\cite{li}}$
suggested that the decays $J/\psi, \psi' \ra \rho \pi$ proceed 
through the intrinsic charm 
component of the $\rho$ wavefunction.  They argue that the 
$c \bar{c}$ pair in the $\mid u \bar{d} c \bar{c}>$ Fock state of the
$\rho^+$ has a nodeless wavefunction which gives it a larger overlap 
with $J/\psi$ than $\psi'$.  This model dramatically challenges the
assumption that charmonium states necessarily decay via intermediate
gluons to exclusive low mass hadrons, deploying the analogy with the $s
\bar{s} (\phi)$ case found at LEAR where there is evidence in $p
\bar{p}$ annihilation at rest of large violation of the OZI rule through
gluon intermediary. However this paper is not yet characterized by
experimentally checkable numbers for the charmonium system.
On the subject of nodes in 
wave-functions, Pinsky$^{\cite{li}}$ had proposed 
several years back that there
is a node in the radial wave function for $\psi'$, 
but not for $J/\psi$, and that this node
makes $\psi' \ra \rho \pi$ a hindered Ml transition like 
$J/\psi \ra \gamma \eta_c$.  However the BES measurements$^{\cite{rad}}$
$\psi' \ra \gamma \eta'$ leads to the ratio
\begin{equation}
{\cal{Q}}_{\gamma \eta'} = \frac{B (\psi \mbox{(2S)} \ra \gamma \eta')}
{B(J/\psi \ra \gamma \eta ')} = \mbox{0.036} \pm \mbox{0.009}
\end{equation}
whereas Pinsky$^{\cite{li}}$ relates the process $\psi' \ra \gamma
\eta'$ to the hindered Ml transition $\psi' \ra \gamma \eta_c$
and predicts $\cal{Q}_{\gamma \eta'}$ = 0.002, which is well below the 
BES measured value.  Finally there is the model of Chen and 
Braaten$^{\cite{li}}$ in which the $c \bar{c}$ pair is in a color octet 
$^3S_1$ state for $J/\psi$.

     Hou and Soni$^{\cite{hou1}}$ developed an earlier suggestion of Freund
and Nambu$^{\cite{hou1}}$ that the decay $J/\psi \ra \rho \pi$ 
is enhanced by the mixing of the 
$J/\psi$ with a glueball $\cal{O}$ that decays to
$\rho \pi$. Brodsky, Lepage, and Tuan$^{\cite{hou1}}$ 
(henceforth referred to as BLT) emphasized that 
$J/\psi \ra \rho \pi$ violates the helicity selection rules of 
perturbative QCD, and argued that the data
requires the glueball $\cal{O}$ to be fairly narrow and nearly
degenerate with the $J/\psi$. It was
however long recognized$^{\cite{brodsky1}}$ that the BLT 
model had a `fly in the ointment' in the known
isospin violating decay $J/\psi \ra \pi^0 \omega$.  
We shall discuss below this mode in some detail.  Present data from
BES constrains the mass and width of the glueball
to the range $\mid m_{\cal{O}} - m_{J/\psi}\mid < 80$
MeV and 4 MeV $< \Gamma_{\cal{O}} < 50$ MeV$^{\cite{hou2,bes}}$.  As
stressed by Chen and Braaten$^{\cite{li}}$
this mass is perhaps 700 MeV (or more) lighter than the lightest 
$J^{PC} = 1^{--}$ glueball observed in
lattice simulation of QCD without dynamical quarks (the ``quenched"
approximation)$^{\cite{bali}}$.
However, lattice calculations for the heavier glueball prediction may be less 
reliable than those for the low mass $J^{PC} = 0^{++}$ 
gluonium state,  since as acknowledged by Bali et al.$^{[8]}$ lattice
studies on the vector glueball are scarce and inconclusive, mainly
because of the difficulties in constructing the corresponding lattice operators.
Indeed
recent QCD sum rule work also find $m (1^{--}) \cong 3.1$ GeV(!), and a
3 GeV $1^{--}$ glueball is quite reasonable according to Brodsky$^{\cite{brodsky2}}$.

     A major problem for the Omicron $\cal{O}$ gluonium 
explanation of the $\rho \pi$ puzzle,
as well as other serious contending models which have an underlying
assumption that hadron helicity conservation HHC$^{\cite{brodsky3}}$
holds at both $J/\psi$ and $\psi '$ (included here are the explanations
of Brodsky and Karliner$^{\cite{li}}$ as well as Chen and Braaten$^{\cite{li}}$
which differ in details about the importance of $D \overline{D}$ channel
and end point form factor$^{\cite{brodsky2}}$), is the known relatively large decay rate for 
$J/\psi \ra \pi^0 \omega ^{\cite{brodsky1}}$, 
the PDG branching ratio$^{\cite{barnett}}$ is $(4.2 \pm 0.6) \times 10^{-4}$, approximately three times larger than the $J/\psi \ra \pi^+ \pi^-$ 
rate.  Both of these
I=1 decays are at first sight presumed to be due to electromagnetic decay (via a highly
virtual $\gamma (Q^2), Q^2 \gg 0)$ or $gg\gamma$, 
where the helicity properties of their amplitudes are
identical to the strong decay $J/\psi \ra ggg \ra \rho \pi ^{\cite{brodsky2}}$, 
and thus should satisfy the requirements
of PQCD helicity conservation$^{\cite{brodsky3}}$. But there is no sign of 
suppression due to hadron
helicity conservation for $J/\psi \ra \omega \pi^0! \ \ $ One possibility is
that there are
additional $q\bar{q}g \ I = 1$
resonances in the 3 GeV mass range which contributes to the 
$\omega \pi^0$ channel (indeed the
Omicron could be the I = 0 member of a hybrid $q \bar{q}g$ 
state at 3 GeV).  But this would destroy the elegance of 
Hou's argument$^{\cite{hou2}}$ on {\underline{why}} 
a gluonium $1^{--}$ should be at 3 GeV and in
any case is a contrived band aid solution.  
Moshe Kugler$^{\cite{kugler}}$ 
suggested that the large
then known $J/\psi \ra \omega \pi^0$ mode could be 
attributed to $\rho - \omega$ mixing.  However$^{\cite{chao}}$ the 
$\rho- \omega$ mixing 
model cannot give the large rate of $J/\psi \ra \omega \pi^0$
through $J/\psi \ra \rho^0 \pi^0 \ra \omega \pi^0$,
because experimentally the rate of $J/\psi \ra \omega \pi^0$ 
is as large as 0.1 times the rate of 
$J/\psi \ra \rho^{0} \pi^{0}{^{\cite{barnett}}}$, 
whereas the effect of $\rho-\omega$ mixing is only of order $10^{- 2}$ 
(note that, e.g. the branching ratio of $\omega \ra \pi^+\pi^-$ 
is about 2\%).  Thus Brodsky presciently suggested even in 1989$^{\cite{brodsky1}}$, 
that in any event it will be {\underline{very important}} to compare these 
branching ratios for $\pi^+\pi^-$, and $\omega^0 \pi^0$
at the $\psi'$ and off resonance.     

One expects the $J/\psi$ and $\psi'$  mesons to decay to hadrons via 
3g or for $\omega \pi^0$ via $\gamma$ or $gg\gamma$ . 
In either case the decay proceeds via $\mid \psi (0) \mid^2$, 
where $\psi (0)$ is the wave function at the origin in the non 
relativistic quark model for $c \bar{c}$.  
Thus it is reasonable to expect on the basis of 
perturbative QCD that for any exclusive hadronic final state h 
(including $\omega \pi^0)$, the branching 
fractions scale like the branching fractions in $e^+e^-$, to wit

\begin{equation}
S=0.14B(J/\psi \ra h) /B(\psi' \ra h) \cong S_{e^{+}e} \cong 1.
\label{fraction}
\end{equation}
 Hence following up on Brodsky's concern$^{\cite{brodsky1}}$, 
it was argued$^{\cite{tuan}}$ that an intriguing 
experimental measurement, if hadron helicity 
conservation HHC Theorem$^{\cite{brodsky3}}$ is \underline{not} applicable 
at both $J/\psi - \psi'$ 
mass range, is to measure $\psi' \ra \pi^0 \omega$ at the level given 
by the 14\% rule of Eq.(\ref{fraction}), i.e, 
0.14 $\times B (J/\psi \ra \omega \pi^0) \sim 0.6 \times 10^{-4}$.  
The BES data$^{\cite{olsen}}$ for $B(\psi' \ra
\omega \pi^0)$ is
$(0.40 \pm 0.10 \pm 0.06) \times 10^{-4}$ and hence taking into account
experimental errors,
the $\omega \pi^0$ decay is consistent with the 14\% rule (2).  
It would appear that only one final test is needed
to shut out the relevance of HHC$^{\cite{brodsky3}}$ for the $J/\psi/\psi'$ 
mass region, and this is to measure the $\psi' \ra \pi^+ \pi^-$
rate, which could be about three times smaller than $\psi' \ra \omega
\pi^0$, if we believe in the analogy with $J/\psi \ra \pi^+ \pi^-$.
Note also the stringent bound on$^{\cite{barnett}}$
 $B(J/\psi \ra \phi \pi^0)$ 
as well as the upper limit for $B
(\psi' \ra \phi \pi^0)$ of $< 0.66 \times 10^{-5}$ (90\% C.L.) in BES
data$^{\cite{olsen}}$ for these companion isospin violating
decays. The situation here can however be 
understood in terms 
of the analysis of Haber \& Perrier$^{\cite{haber}}$, namely the 
reduced branching ratio $\tilde{B} (J/\psi \ra PV) = B(J/\psi \ra PV)/p^3_{V} (p_V$ 
is momentum of vector meson in the center of mass) satisfy
\begin{equation}
\frac{\tilde{B} (J/\psi \ra \pi^0 \phi)}{\tilde{B} (J/\psi \ra \pi^0
\omega)} =
\left [ \frac{1-(2)^{\frac{1}{2}} \tan \theta_V}{\tan \theta_V +
(2)^{\frac{1}{2}}}
\right ]
\end{equation}
for nonet symmetry. If $\phi-\omega$ are assumed to be 
ideally mixed as well 
$[\tan \theta_V = (1/2)^{\frac{1}{2}}]$, then $B(J/\psi \ra \pi^0
\phi)$ vanishes.  
A similar situation can be anticipated for $B(\psi ' \ra \pi^0 \phi)$.

We must not however rush to judgment on the demise of HHC for 
$J/\psi/\psi'$.  First, the experimental `support' for HHC in $J/\psi
\ra
\omega \pi^0$ by Baltrusaistis et al.$^{\cite{baltrusaitis}}$ is a
qualitative one based on its steeper drop in form factor from $q^2 = 0$
to
$q^2 = m_{J/\psi}^2$ when compared with analogous $\pi \pi$ form factor
measured
in $J/\psi \ra \pi^+ \pi^-$.
Pakvasa$^{\cite{pakvasa}}$ pointed out that it is known from the Sutherland 
Theorem based on PCAC that the isospin violating decay $\eta^0 \ra
\pi^+ \pi^- \pi^0$ vanishes for an electromagnetic intermediary, 
but rather is due to $m_d - m_u$ current quark mass difference effect. In the 
present context $J/\psi \ra \pi^0 \omega$ might have a 
dominant decay amplitude $\sim (m_d -m_u)/m_c \cong \alpha/\pi ^{\cite{brodsky2}}$
from this `strong' effect not covered by HHC, and the 
electromagnetic intermediary
$\gamma, \ gg \ \gamma$ governed by HHC to be 
$(\alpha/\pi) \times$ [helicity
suppression].  Perhaps we are seeing the 
former mechanism at work in the decays $J/\psi, \psi' \ra \omega \pi^0$.
Our understanding of $m_u-m_d$ effect  in charmonium decays is however
clearly inadequate still.  Though like Brodsky and Karliner$^{\cite{li}}$
(in a different context) we have dispensed with the $\gamma^*$, gg
$\gamma,$ 3g contributions in $\psi'$ and $J/\psi$ decays to $\omega
\pi^0$, it remains a mystery why for instance isospin violating $\psi'
\ra \rho \pi$ decay has {\underline{not}} been seen through this
mechanism$^{\cite{brodsky2}}$.  Indeed the limit$^{\cite{olsen}}$
on $\psi' \ra \omega \eta < 0.26 \times 10^{-4}$ (90\% C.L.) when
compared with $\psi' \ra \omega \pi^0$ rate may pose another example of
this difficulty.
Hence the present note is to be regarded as a
{\underline{stimulus}} for further theoretical study on the needed
physical idea to complete our understanding.  Setting aside this
obstacle, it is amusing that a consistent picture can 
be sketched for all presently known
$\psi' \ra PV$ decays$^{\cite{olsen}}$ in which 
HHC is valid at both $J/\psi, \psi'$ mass scales for 
the strong 3g intermediary
decays, based on the BLT model$^{\cite{hou1}}$ 
supplemented by recent work$^{\cite{hou2}}$.

      Seiden, Sadrozinski, and Haber$^{\cite{seiden}}$ presented a general phenomenological 
parametrization of amplitudes for $J/\psi \ra P+V$ in their Table IV.
We adapt their analysis for $\psi' \ra P+V$ using their notation.  Hence 
g = strong singly disconnected SOZI amplitude with 3 
gluons exchanged, e is the
``electromagnetic" amplitude which may actually arise from 
the $(m_d-m_u)$ effect discussed
above but expected to be comparable to the usual electromagnetic
strength, r the doubly
disconnected DOZI suppression factor breaking nonet symmetry.  The SU(3) 
violation is accounted for by a factor $(1-s)$ for every strange quark 
contributing to g, a factor $(1-s_p)$ for a strange pseudoscalar contributing 
to r, a factor $(1-s_v)$ for a strange vector contributing to r, and a factor 
$(1-s_e)$, if relevant, for a strange quark contributing to e. One crucial 
ingredient to the analysis is the quark content of the $\eta$ and $\eta'$.  
We shall argue below that the BES data$^{\cite{olsen}}$ are in fact consistent 
with the assumption that the dominant part of the $\eta$ and $\eta'$ 
wavefunctions consists solely of $u\bar{u}, d\bar{d}$, and $s\bar{s}$.
It has been known for sometime$^{\cite{chao1}}$, that the $c \bar{c}$
contribution to $\eta$ and $\eta'$ is miniscule.  Following$^{\cite{seiden}}$, we write
\begin{eqnarray}
\eta & = & X_\eta \mid u \bar{u} + d \bar{d}   > / 2 ^{1/2} + Y_\eta \mid s
\bar{s} > \\ \nonumber
\eta' & = & X_\eta' \mid u \bar{u} + d \bar{d} > / 2^{1/2} + Y_{\eta'} \mid s
\bar{s} >
\end{eqnarray}
and take their approximate values for the X's and Y's 
extracted from the two photon width
of the $\eta, \eta'$, to wit
\begin{eqnarray}
%\begin{equation}
X_\eta  & = & 0.8, Y_\eta = -0.6 \\ \nonumber
X_{\eta'} & = & 0.6, Y_{\eta'} = 0.8   
%\label{towit}
\end{eqnarray}

For $J/\psi \ra P+V$, s is small $\sim$ 10 - 20\% of g and r $\sim 0.15$; 
we shall assume the same
for $\psi' \ra P+V$. Indeed r, s, $s_p$, and $s_v$ 
are all small numbers$^{\cite{seiden}}$ for $J/\psi \ra P+V$. It
seems reasonable, given the preliminary nature 
of the BES data$^{\cite{olsen}}$ for $\psi' \ra P+V$, to make
the same small number assumption here.  Where needed, and for 
definiteness, we take
\begin{equation}
s_e = s_v = s_p = s = 0.15
\label{take}
\end{equation}
in rough accord with expectations in $J/\psi \ra P+V^{\cite{seiden}}$.

The analysis proceeds as follows.  First $\psi' \ra \rho \pi$ 
has not been seen, so \underline{each} of $\rho^+\pi^-, \rho^0 \pi^0, \rho^- \pi^+$
with amplitude g + e, must satisfy
\begin{equation}
g + e \cong 0, \mbox{or} \ g \cong -e.
\label{satisfy}
\end{equation}
This leads to the remarkable conclusion that $\psi' \ra \rho \pi$, 
via 3 gluon strong (3g) decay is
suppressed to the usual electromagnetic transition level $\sim
\alpha/\pi$ in amplitude strength.
Hence Brodsky-Lepage HHC Theorem$^{\cite{brodsky3}}$ appears to work for $\psi' \ra
\rho \pi$ strong (3g) decay, and
furthermore Eq.(\ref{satisfy}) gives us a concrete measure of the magnitude of such helicity 
suppression. If we ignore small s, $s_e$  contributions, then for $\psi'
\ra K^*\overline{K}$ \underline{each} of $K^{*+} K^-, K^{*-}K^+$ has amplitude
\begin{equation}     
g(1 - s) + e(1 + s_e) \cong g + e \cong 0 
\end{equation}
which is of course consistent with existing
$\psi' \ra K^{*+}K^-$ upper limit$^{\cite{olsen}}$. 
For $K^{*0} \overline{K}^0, \overline{K}^{*0} 
K^0$ decay of $\psi'$, the amplitude for each is
 \begin{equation}
g(1-s) - e (2-s_e) \cong -3e.
\end{equation}
This is consistent with {\underline{finite}} BES number for $\psi' \ra
K^{*0} \overline{K}^0$, B $(\psi' \ra K^{*0}\overline{K}^0) =(0.84 \pm
0.24 \pm 0.16) \times 10^{-4}$ $^{\cite{olsen}}$.
The central value is a little higher than the branching ratio 
for $\psi' \ra \pi^0 \omega$ which has also
amplitude 3e.  But within errors the two numbers are
consistent; besides Eq.~(\ref{satisfy}) is only
approximate $\mid g \mid$ may still be slightly larger than 
$\mid e \mid$ leading ultimately to a finite
branching ratio for $\psi' \ra \rho \pi$.  To put in some numbers and
give $\psi' \ra K^{*+} K^-$ a target estimate
to shoot for, let us take $s = s_e$ = 0.15
for SU(3) breaking effects and use the expressions 
given on left hand side of (9) with the same phase space for charged and neutral members of
$K^*\overline{K}$. We have [using the central value for $B( \psi'\ra K^{*0} \overline{K}^0)$].
%from Table~\ref{mesons})].
\begin{equation}
          B(\psi ' \ra K^{*+}K^-) = 1.008 \times 10^{-6}
\label{value}
\end{equation}
Assuming g + e = 0, and ignoring the small $s_p$ contribution (with phase space factor
$p^3_{\omega \eta}/p^3_{\omega \eta'}$ = 1.182), 
we have from Seiden et al.$^{\cite{seiden}}$ Table IV adapted
to $\psi'$ and Eq. (5) 
using the central value for the
branching ratio $B(\psi' \ra \omega \eta'$)$^{\cite{olsen}}$ = $(7.9 \pm 3.6
\pm 1.5) \times 10^{-5}$
 \begin{equation}
           B( \psi' \ra \omega \eta) = (0.097)\times  10^{-4} \cdot
\label{branching}
\end{equation}
This is still consistent with the BES upper limit$^{\cite{olsen}}$ for $B (\psi' \ra
\omega \eta) < 0.26 \times 10^{-4}$.  We shall not make estimates 
for $\psi' \ra \phi \eta'$ from the known experimental
number for $\psi' \ra \phi \eta$, because as seen from 
Seiden et al.$^{\cite{seiden}}$ the theoretical expressions here has
the full gamut of parameters $(g, s, s_e, s_v, r, e, s_p, X, Y).$ It would be easy but not
particularly illuminating to obtain a branching fraction for $\psi' \ra
\phi \eta'$ compatible with
the current experimental limit by adjusting the many parameters available in this case.
The `electromagnetic' transitions $\psi' \ra \rho^0 \eta, \rho^0 \eta'$,
where amplitudes$^{\cite{seiden}}$ are respectively $3 e X_\eta$ and $3e
X_{\eta'}$, can be related to the known $\psi' \ra \omega \pi^0$ rate
with amplitude 3e using Eq. (5).
Taking the phase space factors into account $[p^3_{\rho o
\eta}/p^3_{\omega \pi^0} =
0.935$ and $p^3_{\rho^0 \eta'}/p^3_{\omega \pi^0} =0.793]$ 
and again using the central experimental value for $\psi'\ra \omega \pi^0$
we have
\begin{equation}
B(\psi' \ra \rho^0 \eta) = 0.239 \times 10^{-4}, B(\psi' \ra \rho^0
\eta') = 0.115 \times 10^{-4} \cdot
\label{central}
\end{equation}
These results are quite consistent with BES values$^{\cite{olsen}}$ that
$B(\psi' \ra \rho^0 \eta) \ = (0.21 \pm 0.11 \pm0.05) \times 10^{-4}$
and
$B(\psi' \ra \rho^0 \eta') < 0.3 \times 10^{-4}$ (90\% C.L.).
The prediction of
Eq.~(\ref{value}) may require a larger sample of $\psi'$ 
than the 3.8 million currently available.  This
may be feasible with the construction of MDC III and a future tau-charm factory.  We have
not discussed the $\psi' \ra \gamma \eta, \gamma \eta'$
modes of BES$^{\cite{olsen}}$ because even though 
of the VP type, they involve an external $\gamma$ which is not covered by the 
hadron helicity conservation HHC theorem of 
Brodsky-Lepage$^{\cite{brodsky3}}$ central to our discussion.

     To summarize, amongst the models proposed to explain the strong 
suppression in the $\rho \pi$,
and $K^*\overline{K}$ channel with some claim to a quantitative basis, 
the $J/\psi - {\cal{O}}$ mixing model$^{\cite{hou1,hou2}}$ appears to
survive the latest results, \underline{except} for isospin violating PV modes. 
This model requires the ${\cal{O}}$ mass to be quite near the
 $J/\psi$ and to have a substantial decay width to two-body VP meson states, as particularly
 emphasized by BLT$^{\cite{hou1}}$.  Although searches for this state in an
$e^+e^- \ra \rho \pi$ scan near the
 $J/\psi$ mass$^{\cite{bes}}$ and in $\psi' \ra \pi^+ \pi^- \rho \pi$ 
decays have been negative, they have not provided very
severe restrictions on the properties of the ${\cal{O}}$. A better ${\cal{O}}$
search strategy may be needed.
However, we have already seen from an analysis of the preliminary 
BES $\psi'$ data for PV,
that within this model framework, significant insights into hadronic physics have been
obtained, in particular the strength of the HHC 
suppression$^{\cite{brodsky3}}$ to electromagnetic
level at $\psi'$ and a new departure in the understanding of isospin
violating decays of $J/\psi, \psi' \ra \omega \pi^0$, which however
\underline{demands further study at a fundamental level}.

{\underline{Remarks}}
(i) The $\psi' \ra$ VT final states have been measured$^{\cite{bes1}}$ and found to be
suppressed by a factor of at least three compared to the expectations of
Eq. (2).  Since hadronic VT decays conserve HHC$^{\cite{gu}}$, some other
mechanism must be responsible for this (possibly mild) suppression in
the BLT model$^{\cite{hou1,hou2}}$ and that of Chen and Braaten$^{\cite{li}}$. 
This pattern can be explained by also taking into account the
orbital-angular-momentum selection rule for exclusive amplitudes in
perturbative QCD~$^{\cite{chernyak}}$ and is under study by Chen and Braaten.  
The Brodsky and Karliner model$^{\cite{li,brodsky2}}$ can account for comparable
suppression of both VT and VP modes since the tensor mesons here could
have an appreciable intrinsic charm content.
There is however a general belief that intrinsic charm content of low
lying V, T mesons is \underline{small}$^{\cite{suzuki}}$.
 (ii) Though the models of BLT$^{\cite{hou1,hou2}}$ as described here and 
Chen and Braaten$^{\cite{li}}$ have predictions for $\psi' \ra \omega \pi$, 
in order of magnitude agreement with existing data$^{\cite{olsen}}$, they are 
based on the phenomenological treatment of Ref. [18] rather than a
{\underline{fundamental}} understanding of this isospin violating mode
in the context of HHC, which remains a shared concern with the Brodsky
and Karliner model$^{\cite{li}}$ also. Furthermore, the need for Chen and 
Braaten$^{\cite{li}}$ to invoke $c\overline{c}$ color octet
$^3S_1$ contribution to $J/\psi$ could be a dubious assumption, since
Kroll$^{\cite{kroll}}$
has arguments on why, for some exclusive $J/\psi$ decays at least, the
color octet contribution is an O$(v^2/c^2)$ effect and  hence
negligible
$(< 12\%$ or so).  Mahiko Suzuki$^{\cite{li}}$ in his very thorough paper
demonstrated that the existing data$^{\cite{barnett}}$ on 
$J/\psi \ra \omega \pi^0, \rho \pi, K^* \overline{K}$  assuming that these 
decays are mediated by conventional $\gamma^*$ and $3g$ respectively, would 
lead to the uncomfortable situation of a large relative phase nearly $90^0$ 
between the three-gluon and the one-photon amplitude.  Hence he is pessimistic 
about extracting meaningful information on B-meson decay (presumably at the
future B-factory facilities) on the parameters of the fundamental
interactions here.   Indeed he has gone further$^{\cite{suzuki1}}$ to show
that the large phase problem (and possible resolution of the puzzle) may
rest with measurement of $e^+e^- \ra \gamma \ra \pi^+ \pi^-$ and 
$e^+e^- \ra \gamma \ra K^+K^-$ off the $J/\psi$ without recourse to
theory.  However construction of MDC III and a future tau-charm factory
may be needed to obtain this decisive measurement.
We agree with Suzuki that a high priority of future
$e^+e^-$ facilities, e.g. the $5\times 10^7 J/\psi$ events expected from
the BES  upgrade/MDC III, is the high precision
{\underline{remeasurement}} of the $J/\psi$ decay branchings,
{\underline{particularly for $\rho \pi, K^* \overline{K},$ and
$\omega \pi^0$.}} (iii) Though it is hoped that eventually lattice gauge
work would provide accurate prediction of the width of gluonium states
like the Omicron, we note that the limit currently$^{\cite{hou2,bes}}$ set on
Omicron width $\Gamma < 50$ MeV is already a severe one.  At round 3.1
GeV a 3g glueball cannot be too  narrow$^{\cite{brodsky2}}$.  Hence, we urge 
renewed effort at BES upgrade/MDC III to continue the search for the Omicron
vector glueball by a scan of the $J/\psi$ resonance.  Remember that the
HHC theorem$^{\cite{brodsky3}}$ was developed with very minimal assumptions
from perturbative QCD.  If it should prove to be invalid at the $J/\psi$
and $\psi'$ mass range, the validity of Eq. (2) can also be questioned.

I would like to thank S.J. Brodsky, S. Pakvasa, S.L. Olsen, W.S. Hou,
Y.F. Gu, J.L. Rosner, and M. Suzuki for helpful discussions and
communications; I also wish to acknowledge the support in 
part by the U.S. Department of Energy under Grant DE-FG-03-94ER40833 at
the University of Hawaii-Manoa.

%\section*{References}

\end{document}